\documentclass[runningheads]{llncs}
\usepackage[T1]{fontenc}

\usepackage[utf8]{inputenc}

\usepackage{amsthm}
\theoremstyle{plain}
\newtheorem{thm}{Theorem}
\theoremstyle{definition}
\newtheorem{defn}[thm]{Definition} 

\usepackage[linesnumbered,ruled,vlined]{algorithm2e}
\usepackage{enumitem}
\usepackage{multirow}
\usepackage{multicol}
\usepackage{booktabs}
\usepackage{float} 
\usepackage{makecell}
\usepackage{graphicx}
\usepackage[normalem]{ulem}
\usepackage[font=footnotesize]{subfig}
\usepackage[misc,geometry]{ifsym}
\usepackage{amsmath}
\usepackage{xcolor}

\begin{document}
\title{Re-thinking Human Activity Recognition with Hierarchy-aware Label Relationship Modeling}
%
%

\author{Jingwei Zuo \orcidID{0000-0002-3251-6939} (\Letter) \and Hakim Hacid \orcidID{0000-0003-2265-9343}}
\authorrunning{J. Zuo \and H. Hacid}
\titlerunning{H-HAR: Re-thinking HAR by Modeling Hierarchical Label Relationship }
%
\institute{Technology Innovation Intitute, Abu Dhabi, UAE\\
\email{\{jingwei.zuo, hakim.hacid\}@tii.ae}}
\maketitle              
\begin{abstract}
  Human Activity Recognition (HAR) has been studied for decades, from data collection, learning models, to post-processing and result interpretations. However, the inherent hierarchy in the activities remains relatively under-explored, despite its significant impact on model performance and interpretation. 
  In this paper, we propose H-HAR, by rethinking the HAR tasks from a fresh perspective by delving into their intricate global label relationships. Rather than building multiple classifiers separately for multi-layered activities, we explore the efficacy of a flat model enhanced with graph-based label relationship modeling. 
  Being hierarchy-aware, the graph-based label modeling enhances the fundamental HAR model, by incorporating intricate label relationships into the model. 
  %
  We validate the proposal with a multi-label classifier on complex human activity data. The results highlight the advantages of the proposal, which can be vertically integrated into advanced HAR models to further enhance their performances.
  
\end{abstract}

\section{Introduction} \label{sec:intro}


Human activity recognition (HAR) has gained, in recent years, a great interest from both the research community and industry players. The activity data can be collected from multiple data sources~\cite{zuo2023handling}, such as GPS trajectories, web browsing records, smart sensors, etc. Among which, the human physical activity with wearable sensors are widely studied~\cite{zuo2021smate}, which are represented by multivariate time series (MTS). 
When studying the HAR tasks, the focus is typically on the complexity of the activity data itself, such as dealing with complex MTS formats, processing noisy data, or identifying multiple sequential actions within a single activity. 
Existing research often explores a flat classifier approach that learns complex activity data. However, the complexity in the label relationships between activity class labels is usually overlooked. 

\begin{figure}[ht]
    \centering
    \includegraphics[width=0.7\linewidth]{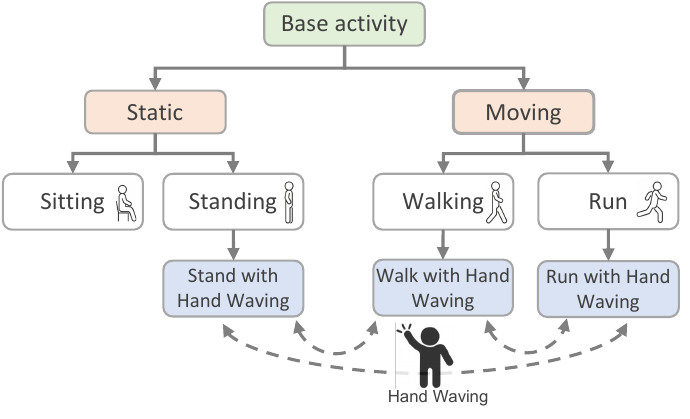}
    \caption{The label structure in physical activities, including predefined relationships (solid lines) and implicit relationships (dashed lines).
    }
    \label{fig:har_graph_feature}
\end{figure}

As shown in Figure~\ref{fig:har_graph_feature}, an inner label structure always exists in physical activities, providing rich information for building a reliable HAR classifier. 
%
Recent work \cite{leutheuser2013hierarchical,debache2020lean,zheng2015human,zhang2010activity} consider the hierarchy features between human activities, and have proved that a hierarchy-aware model shows better performance in terms of model's reliability and efficiency.
However, these methods employ a straightforward top-down approach, constructing individual classifiers at each hierarchical level. Classifiers in lower layers are based on predictions from upper-layer classifiers, a local-based process that introduces several limitations:
i) Multiple classifiers need to be built at each level, leading to escalating complexity within the hierarchical structure; 
ii) Classifiers at each level focus on relationships within the same layer, under a common parent node, ignoring the broader, global relationships between the cross-layer activities; 
iii) Classifiers at lower levels rely on the predictions from upper layers as their training annotations, leading to larger accumulative errors in a deeper hierarchical structure. 
%
Moreover, as depicted in Figure \ref{fig:har_graph_feature}, the relationships between class labels depend not only on the pre-defined hierarchical structure, but also on the implicit, hidden links between certain activities. Therefore, relying on a pre-defined label structure risks overlooking these vital relationships between activities.

To tackle the above-mentioned challenges, we propose H-HAR, a Hierarchy-aware Model for HAR tasks. Instead of building multiple local HAR classifiers, we learn a flat model to process the activities in a global manner. Precisely, we embed and project the label hierarchy into the representation space of the data. The label and data hierarchy will be carefully aligned, allowing the model to benefit the rich information from both data and the hierarchy features. 
To build the representation space, a graph-based label encoder and an activity data encoder are proposed: the label encoder learns complex label embeddings by combining a predefined label hierarchy and a learnable graph structure; whereas the data encoder builds data embeddings of input activities. The aligned label-data embeddings are learned via a supervised contrastive loss~\cite{khosla_supervised_2020}, considering inter-class and intra-class embeddings. Concretely, the nearby neighbors in the hierarchy will stay close to each other in the representation space, while distant nodes will keep far away from each other. A multi-label classifier is jointly built over the representation space.

To summarize, our contributions in this paper are as follows:

\begin{itemize}
    \item \textbf{Label relationship modeling}: we propose a label encoder that automatically learns the label relationships without a predefined label structure. 
    \item \textbf{Embeddable label encoder with scalability}: the label encoder can be seamlessly integrated into other HAR models to learn better representations.
    \item \textbf{Label-data semantic alignment}: we align the label and data semantics in the representation space, allowing building class-separable data embeddings.
    \item \textbf{Joint embedding \& multi-label classifier optimization}: we jointly optimize the embedding space and classifier, providing reliable performance. 
    
\end{itemize}

\section{Related work} \label{sec:related_work}
In this section, we describe the most related work of our proposal in Human Activity Recognition (HAR) tasks and Hierarchical Label Modeling.

\subsection{Human Activity Recognition (HAR)}
Human Activity Recognition (HAR) is a largely investigated domain. In our context, we consider human physical activities, with data acquired easily from smart sensors.  
The sensor-based activity data is generally represented by Multivariate Time Series (MTS)~\cite{zuo2021smate}. As a classification task, the HAR can be based on various feature extractors (i.e., feature representations) and classifiers. For instance, one can use handcrafted statistical features~\cite{zuo2023handling,zuo2024magneto} to feed any classifiers, which is easy-to-deploy and requires linear processing time.
Other work in MTS Classification domain, where researchers aim to build general ML models covering various applications~\cite{zuo2021smate}, including HAR tasks. For instance, Shapelet features~\cite{zuo2019exploring} with a kNN classifier, or end-to-end neural network models~\cite{zuo2021smate}.

However, these approaches usually focus on handling the complex activity data, overlooking the complex label relationships, that provides rich information for building reliable feature representations.  


\subsection{Hierarchical Label Modeling}
Inherently, there exists a hierarchical label structure in human activities. The label hierarchy, as shown in Figure~\ref{fig:har_graph_feature}, can be considered as a tree-based structure. It allows enriching the data embeddings and forge a robust representation space to learn class-separable embeddings.
Rarely investigated in HAR tasks, the label modeling is usually studied in Natural Language Processing (NLP) applications, where the hierarchy features widely exist in the semantic labels~\cite{zhou2020hierarchy}. A typical example is Hierarchical Text Classification (HTC)~\cite{zhang2016embedding}, for which a sentence can be tagged with different labels. A multi-label classifier can be built over the text representations, which can be improved by considering the label relationships. 

The hierarchical label modeling in previous studies\cite{zhou2020hierarchy} can be either local or global approaches. 
Local approaches~\cite{banerjee2019hierarchical,dumais2000hierarchical,shimura2018hft} build multiple classifiers at each hierarchical level. However, they basically ignore the rich structural interactions between nodes at a global scale, i.e., the activities can share common patterns even though they do not share the same parent nodes. For instance, in Figure~\ref{fig:har_graph_feature}, \textit{still with hand waving} and \textit{walking with hand waving} are two activities under \textit{still} and \textit{walking}. Though having different parent nodes, they share the same action of \textit{hand waving}. In consequence, the local modeling approaches only capture limited interactions between neighboring activity nodes in the hierarchy structure.
Previous HAR models \cite{leutheuser2013hierarchical,debache2020lean,zheng2015human,zhang2010activity} usually model the hierarchy in this manner, i.e., training multiple classifiers at different hierarchical levels.

As for global approaches~\cite{shimura2018hft,zhou2020hierarchy,chen2021hierarchy,wang2022incorporating}, they build a flat-label classifier for all classes. Therefore, how to integrate the hierarchy information into the model becomes the research focus of the recent studies, i.e., building a hierarchy-aware flat-label classifier.
Various work has studied the joint modeling of label and data embeddings in HTC tasks. For instance, authors in \cite{zhang2016embedding,tonioni2019domain} designed a generalized triplet loss with hierarchy-aware margin, which allows differentiating fine and coarse-label classes. With more considerations on the hierarchical information, the work in \cite{zhou2020hierarchy} introduced Prior Hierarchy Information from the training set, which serves to encode the label structures. The label structure can be either encoded by a Bidirectional Tree-LSTM, or a Graph Convolutional Network (GCN). Consequently, the hierarchy-aware label embedding can be combined with text embeddings to feed a multi-label classifier. HiMatch~\cite{chen2021hierarchy} further aligns the text semantics and label semantics, and adopt a similar Triplet loss with a hierarchy-aware margin to accelerate the computation process. 

%
However, the above-mentioned work, both local and global approaches, heavily relies on prior knowledge of the label hierarchy information. In consequence, the implicit, hidden relationships between label nodes are usually ignored, leading to a less optimal modeling of the label relationships.
\section{Problem Formulation}\label{sec:problem_formulation}
In this section, we formulate our research problems on HAR with learnable label relationship modeling.
Table \ref{Math_notation} summarizes the notations used in the paper.

\begin{table}[htbp]
\centering
\caption{Notation}\label{Math_notation}
\scalebox{0.9}{
\begin{tabular}{|p{4.8cm}| l |}
\hline
\small
Notation & Description\\
\hline
\hline
$\mathcal{D} = \{X, L\}$ & Activity and label sets\\
$X = \{X_1, ..., X_{N}\}\; or \;\{x_1, ..., x_{n}\} $ & A sequence of activity sets $X_1, ..., X_N$, sample $x_1, ..., x_n$\\
$L = \{l_1, ..., l_{N}\}\; or \;\{l_1, ..., l_{n}\} $ & A sequence of label sets $l_1, ..., l_N$. (\text{Note:} \textit{multi-label for $x_i$})\\
$N, n$ & Number of label sets, number of samples\\
$\mathbf{E}_{L}=\{e_{1}, ..., e_{N}\}$ & Label embeddings\\
$\mathbf{E}_{X}=\{e'_{1}, ..., e'_{N}\}$ & Data embeddings\\ 
$\mathcal{G} = <\mathcal{V}, \mathcal{E}>$ & A graph including the vertex and edge sets \\
$\varphi: \mathcal{X} \to \mathcal{R}^d$ & Feature map function, e.g., a linear layer\\ 
$\Theta$ & Model parameters\\

\hline
\end{tabular}}
\end{table}

\begin{defn}(Hierarchical Human Activity). We denote the Hierarchical Human Activity data as $\mathcal{D} = \{X, L\}$ with a sequence of activity sets $X = \{X_1, ..., X_{N}\}$ and a sequence of label sets $L = \{l_{1}, ..., l_{N}\}$. Each label set $l_{i}$ contains a set of classes, belong to either one or more sub-paths in the hierarchy.

As shown in Figure \ref{fig:har_graph_feature}, the hierarchical class labels can be formulated as a graph structure. Therefore, each label set $l_{i}$ represents the labels passed through the root node to a terminal node, that can be a leaf or a non-leaf node. 
\end{defn}


\begin{defn}(Hierarchical Human Activity Recognition). Given a data set $\mathcal{D} = \{X, L\}$, we aim to learn a multi-label classifier $f$ from $\mathcal{D}$. For an unseen activity $x_{i}$, the classifier $f$ can accurately predict its label set $\hat{l}_{i} = \{\hat{y}_{i}\}^{m}$, where $m$ is the number of labels.
\end{defn}

\begin{defn}(Hierarchical Label Embedding). Given a sequence of label sets $L = \{l_{1}, ..., l_{N}\}$, we aim to learn a set of hierarchy-aware label embeddings $\mathbf{E}_{L}=\{e_{1}, ..., e_{N}\}$, integrating hierarchical features from $L$ for each target instance.  
\end{defn}

Learning hierarchical human activities requires considering not only the features of activity data, i.e., data embeddings, but also relationships (\textit{explicit} and \textit{implicit}) between activities, i.e., label embeddings. 
We aim to learn a representation space $\mathcal{H}$ where the raw activity data are embedded and aligned with learnable label relationships. 
The learning objective is to minimize the classification loss $\theta = min_{\theta \in \Theta} \texttt{L}(f_{\theta}(X, \mathbf{E}_{L}), Y)$. 
\section{Our proposals}\label{sec:our_proposals}

\begin{figure*}[ht]
    \centering
    \includegraphics[width=1\linewidth]{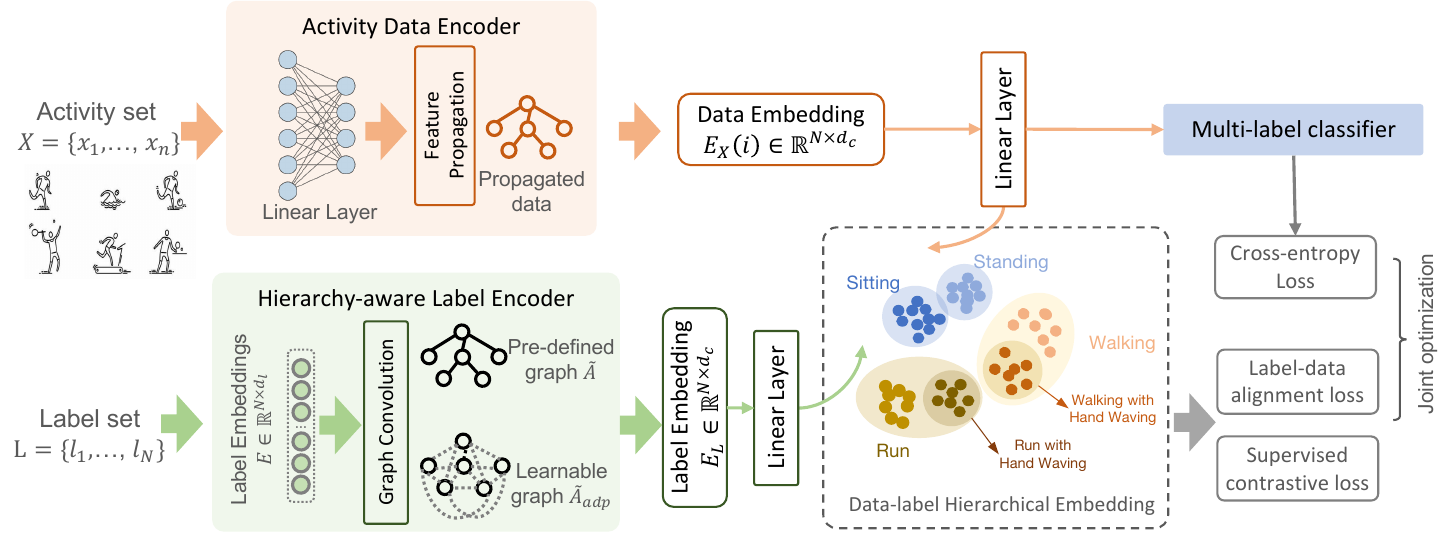}
    \caption{Global system architecture of H-HAR
    }
    \label{fig:system_structure}
\end{figure*}

To handle the aforementioned challenges, we propose H-HAR, a Hierarchy-aware model for HAR tasks. As shown in Figure~\ref{fig:system_structure}, H-HAR relies on a graph-based label encoder and an activity data encoder. 
The graph-based label encoder extracts the complex hierarchical relationships between labels, with a predefined label hierarchy and a learnable graph structure. 
The data encoder simply builds data embeddings of input activities.
By aligning the hierarchical label and data embeddings, H-HAR is able to learn data representations with hierarchy semantics.

\subsection{Hierarchy-Aware Label Encoding}
In the label hierarchy, the nodes under the same parent node share similar patterns. 
Unlike previous studies~\cite{leutheuser2013hierarchical,debache2020lean,zheng2015human,zhang2010activity} considering only child-parent and child-child relationships, we consider global node relationships, coming with more discriminative features. 
%
Due to the intricate global relationships among the label nodes, it is natural for us to represent the label hierarchy as a graph. 
\begin{defn}(Hierarchy as Graph). We define a label graph $\mathcal{G} = <\mathcal{V}, \mathcal{E}>$ to represent the hierarchical structure among the labels, where $\mathcal{V}$ = $\{v_{1}, v_{2},...,v_{N}\}$ denotes the label set with $N$ nodes, $\mathcal{E}$ = $\{(v_{i}, v_{j}) | v_{i} \in \mathcal{V}, v_{j} \in link(i)\}$ indicates the directed edge connections between $v_{i}$ and it's linked nodes. 
\end{defn}
In the label graph $\mathcal{G}$, each activity is regarded as a node, and can be connected or disconnected from others. The graph edges represent the node relationships. 
We should note that the relationships are not fully decided by a pre-defined label hierarchy, i.e., edge connections.
As aforementioned in Figure~\ref{fig:har_graph_feature}, the implicit hidden relationships exist for unconnected nodes in the pre-defined hierarchy. 
Therefore, we propose to learn the implicit node relationships via a learnable hidden graph. The pre-defined and learnable graphs are jointly considered in a Graph Convolution Network (GCN)~\cite{zuo2023graph}, to build hierarchy-aware label embeddings.

\subsubsection{Predefined graph structure}
In GCNs~\cite{kipf2017semi}, the adjacency matrix represents the graph connections or relationships between the nodes. We define $A_{i,j} = \frac{\left|H(v_i)\cap H(v_j)\right|}{\left|H(v_i)\right|}$ as the connection weight between node $v_i$ and $v_j$, where $H(v)$ denotes a set of higher level nodes (i.e., all the parent nodes of $v$). Intuitively, $\left|H(v_i)\cap H(v_j)\right|$ represents the number of shared parent nodes between $v_i$ and $v_j$, $A_{i,j}$ shows the proportion of common ancestors over the node $v_i$. A larger value of  $A_{i,j}$ represents a closer relationship between $v_i$ and $v_j$. 

Let $\tilde{\mathbf{A}} = I + D^{-\frac{1}{2}} A D^{-\frac{1}{2}} \in \mathcal{R}^{N \times N}$ denote the normalized adjacency matrix with self-loops, where $D$ is the degree matrix representing the degree of each vertex in the graph. 
Given a sequence of label set $L = \{l_{1}, ..., l_{N}\}$, we define the intermediate label embeddings $E = e(L) \in \mathcal{R}^{N \times d_{l}}$ as the input signals, where $e$ is the embedding function. Then the label embeddings $\mathbf{E}_L$ integrating the graph structural features is defined as the output of a graph convolution layer~\cite{kipf2017semi}:
\begin{equation}
\label{eq:predefined_graph}
    \mathbf{E}_L=\sigma(\tilde{\mathbf{A}} E \mathbf{W_p}) \in \mathcal{R}^{N \times d_{c}}
\end{equation}
where $\sigma$ is ReLU activation, $\mathbf{W_p} \in \mathcal{R}^{d_{l} \times d_{c}}$ denotes GCN's weight matrix.

\subsubsection{Learnable graph structure}
The hidden interactions allow the model to enrich the label embeddings from a global view (i.e., interacting nodes from different layers and branches).
To capture the implicit connections, as a complement of the predefined graph, we learn a self-adaptive graph, that does not require any prior knowledge and is
learned end-to-end through stochastic gradient descent. 
We initialize two random matrices $E_{1}$, $E_{2}\in \mathcal{R}^{N\times d_{f}}$, representing source and target node embeddings~\cite{wu2019graph}. We define the self-adaptive adjacency matrix as:
\begin{equation}
\label{eq:learnable_graph}
    \mathbf{\Tilde{A}}_{adp} = SoftMax(ReLU(E_{1}E_{2}^{T}))
\end{equation}
$E_{1}E_{2}^{T}$ shows the dependency weights between source/target nodes. \textit{ReLU} serves to filter weak connections. \textit{SoftMax} is used to normalize the adjacency matrix.

With the predefined graph in Equation \ref{eq:predefined_graph}, we re-define the graph layer as:
\begin{equation}
\label{eq:predefined_learnable_graph}
    \mathbf{E}_L=\sigma(\tilde{\mathbf{A}} E \mathbf{W_p}
    + \mathbf{\Tilde{A}}_{adp} E \mathbf{W}_{adp}) \in \mathcal{R}^{N \times d_{c}}
\end{equation}

\subsection{Activity Data Encoding}
Raw physical activity data is usually represented as Multivariate Time Series, i.e., $X$ = $\{x_1, ..., x_{n}\} \in \mathcal{R}^{n \times m \times t}$, where $n, m, t$ represent number of instances, sensors and timestamps. In this paper, we focus on the model's label encoding behavior. Therefore, aligned with pre-processed data of multiple HAR datasets (e.g., DaliAc~\cite{leutheuser2013hierarchical}, mHealth~\cite{misc_mhealth_dataset_319}), we consider $X$ = $\{x_1, ..., x_{n}\} \in \mathcal{R}^{n \times d}$, where $d$ is the input feature dimension. More advanced feature extractors on raw data can be explored and integrated into our framework. This is orthogonal to our work.

Given $X \in \mathcal{R}^{n \times d}$, we define the intermediate data embedding $E_d=e(X)\in \textsc{R}^{n\times d_x}$. Following previous work~\cite{zhou2020hierarchy}, we further introduce a graph-based feature propagation module to encode label hierarchy information. The propagation module first reshapes activity features $E_d$ to align with the graph node input:
\begin{equation}
\label{eq:reshape_graph}
    V = E_{d} W_{res} \in \textsc{R}^{n \times N \times d_c}
\end{equation} 
where $W_{res} \in \textsc{R}^{d_x \times N \times d_c }$. Then the GCNs built in Equation~\ref{eq:predefined_learnable_graph} can be employed to integrate label hierarchical information:
\begin{equation}
\label{eq:feature_propagation}
\mathbf{E}_X=\sigma(\tilde{\mathbf{A}} V \mathbf{W_p}'
    + \mathbf{\Tilde{A}}_{adp} V \mathbf{W}_{adp}') \in \mathcal{R}^{n \times N \times d_{c}}
\end{equation} 
Note that $\tilde{\mathbf{A}}$ and $\mathbf{\Tilde{A}}_{adp}$ are shared graphs between the label and data encoding. 


\subsection{Label-data Joint Embedding Learning}
\subsubsection{Label-data alignment} 
Even though the data embeddings are reshaped to align with the graph node input, there is no explicit matching between data embeddings and label embeddings, that contains rich label relationships. To this end, we jointly built label-data embeddings in the representation space to align data and label semantics. Concretely, we apply the L2 loss between data and label embeddings:
\begin{equation}
\label{eq:joint_l2_loss}
\textsc{L}_{\text{align}}= \sum_{i=1}^{n} \|\varphi_{x}(\mathbf{E}_X(i)) - \varphi_{l}(\mathbf{E}_L)\|^2
\end{equation}
where $\varphi_{x}$ and $\varphi_{l}$ are linear layers to project $\mathbf{E}_X$, $\mathbf{E}_L$ to a common latent space.

\subsubsection{Class-separable Embedding Building} The label-data alignment loss only captures the correlations between activity data and labels, while the label embeddings are not clearly separable. To learn class-separable embeddings, we employ a supervised contrastive loss~\cite{khosla_supervised_2020} 
to the representation space: 
\begin{equation}
\small
\label{eq:cont_loss}
\begin{aligned}
\textsc{L}_{con}\left(\mathbf{E}_X(i), \mathbf{E}_X(j), Y\right) &=Y *\left\|\varphi_{X}\left(\mathbf{E}_X(i)\right)-\varphi_{X}\left(\mathbf{E}_X(j)\right)\right\|^2 \\
&+ (1-Y) *\left\{\max \left(0, \text{m}^2-|| \varphi_{X}\left(\mathbf{E}_X(i)\right)-\varphi_{X}\left(\mathbf{E}_X(j)\right) \|^2\right)\right\}
\end{aligned}
\end{equation}
where $\text{m} >0$ is the margin parameter, $Y=1$ if $l_i= l_j$, otherwise $Y=0$.

\subsubsection{Classification and Joint Optimization}
As shown in Figure~\ref{fig:har_graph_feature}, the hierarchy can be flattened for multi-label classification. The data embedding $\mathbf{E}_X$ is followed by a linear layer and a \text{sigmoid} function to output the probability on label $j$:
\begin{equation}
    p_{i j} = \text{sigmoid}\left(\varphi_{X}\left(\mathbf{E}_X(i)\right)\right)_{j}   
\end{equation}
Therefore, a binary cross-entropy loss is applied:
\begin{equation}
    \textsc{L}_{ce} = \sum_{i=1}^{n} \sum_{j=1}^{N} - y_{i j} \log \left(p_{i j}\right)-\left(1-y_{i j}\right) \log \left(1-p_{i j}\right)  
\end{equation}
where $y_{ij}$ is the ground truth: $y_{ij}=1$ if $x_i$ contains a label $j$, otherwise 0.   

We jointly optimize the model by combining the label-data alignment loss, contrastive loss and cross-entropy loss: 
\begin{equation}
    \textsc{L} = \textsc{L}_{align} + \lambda_{1} \textsc{L}_{con} + \lambda_{2}\textsc{L}_{ce} 
\end{equation}
where $\lambda_{1}$, $\lambda_{2}$ are hyperparameters controlling the weight of the related loss. During inference, we only use the Activity Data Encoder for classification. 
\section{Experiments}
\label{sec:experiments}
In this section, we validate H-HAR with real-life human activity datasets. The experiments were designed to answer the following Research Questions (RQs):
\begin{itemize}[leftmargin=.5in]

    \item[\textbf{RQ 1}] \textit{H-HAR Performance}: How does H-HAR compare to other (hierarchical) models in HAR tasks?
    \item[\textbf{RQ 2}] \textit{Label Encoding Efficiency}: How effective is our graph-based label encoding compared to other label modeling methods in HAR? 
    \item[\textbf{RQ 3}] \textit{Impact of Joint Optimization}: What are the benefits of using multiple objective functions together in improving HAR model performance?

\end{itemize}
\subsection{Experimental Settings}
\subsubsection{Dataset Descriptions}
\begin{figure*}[ht]
    \centering
    \includegraphics[width=1\linewidth]{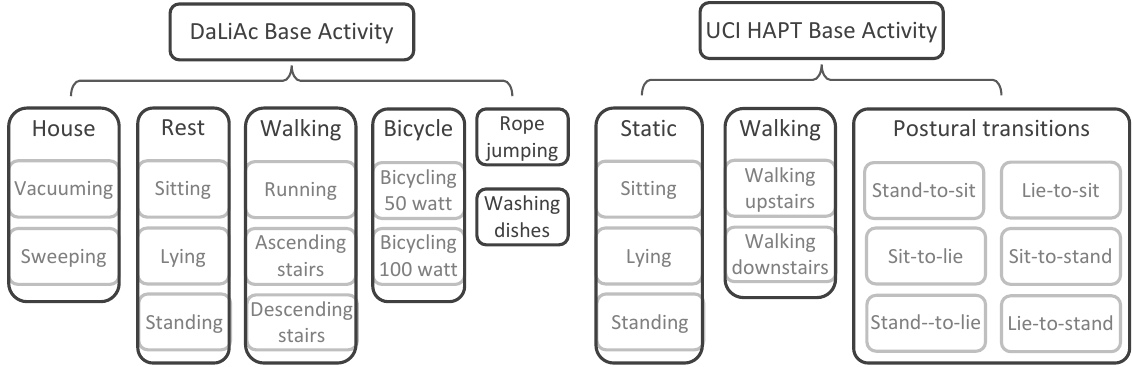}
    \caption{Predefined label hierarchy in DaLiAc and UCI HAPT.
    }
    \label{fig:datasets_hierarchy}
    \vspace{-2em}
\end{figure*}

We choose DaliAc~\cite{leutheuser2013hierarchical} and UCI HAPT~\cite{reyes2016transition} as testing datasets because of their rich label relationships. As shown in Figure~\ref{fig:datasets_hierarchy}, the DaliAc dataset contains 13 activities collected by 19 participants. The UCI HAPT dataset was collected from 30 volunteers, with 6 basic activities and 6 postural transitions. We follow~\cite{debache2020lean,thu2021hihar} as for the data preprocessing and training/testing split. As both datasets are relatively class-balanced, for simplicity, we report the average accuracy of all classes in each dataset.

\vspace{-1em}

\subsubsection{Execution and Parameter Settings} The proposed model is implemented in PyTorch 1.6.0 and is trained using the Adam optimizer in one single Nvidia A100 (40G). We set an adaptive learning rate regarding training epochs, i.e., the learning rate starts from 0.01 and decreases by half every training epoch. We set the balancing weight $\lambda_1=\lambda_2=1$.

\subsubsection{Baselines}
For HAR tasks, much of the existing research adopts local-based approaches. These typically involve constructing multiple classifiers in a top-down manner. They can be essentially simplified to a flat classifier model when not considering the predefined label hierarchy.
Therefore, we selected popular conventional ML models for evaluation, including AdaBoost, kNNs (k=7), SVM, and Multi-layer Perceptron (MLP) with a Softmax activation function. 

It's important to note that while there are numerous advanced models that could potentially yield superior HAR performance, our focus is primarily on examining the impact of label relationship modeling within HAR tasks, rather than identifying the most advanced model architectures.

Additionally, we assessed the performance of these models both with and without considering label hierarchy. For the baseline models, we did not incorporate any predefined hierarchy. In contrast, for H-HAR, we substituted the graph-based label encoding layer with a linear layer. We also extended our evaluation to both single-label and multi-label classification tasks to comprehensively understand the models' behavior.

\subsection{Experimental results}
Table~\ref{table:accuracy_comparison} presents a comparison of the accuracy of various HAR models, both with and without considering label hierarchy (denoted as w/o H. and w/ H. respectively). With advanced label-data embedding learning and joint classifier building, it is not surprising that H-HAR shows superior performances of others. However, the results offer several key insights:
\begin{itemize}
    \item Robustness in Multi-label Classification (\textbf{RQ 1}): While there is a general decline in model performance for multi-label classification tasks, H-HAR exhibits a relatively small decrease compared to other baseline methods. This suggests that H-HAR is robust in differentiating between parent and child node classes.
    \item Improvement with Predefined Label Hierarchy: The introduction of a predefined label hierarchy significantly enhances the performance of baseline models, particularly noted in the SVM on the DaLiAc dataset with an improvement of over 30\%. As illustrated in Figure~\ref{fig:datasets_hierarchy}, building classifiers at each layer effectively reduces the learning complexity by leveraging rich prior label knowledge.
    \item Superiority of Neural Network-Based Approaches: Neural network-based models generally outperform traditional ML models in this context, where the data is straightforward, and the feature space is limited. Exploring more advanced network architectures could further augment the model's performance, which is orthogonal to this work.
\end{itemize}
However, due to a larger parameter space, H-HAR performs less efficient than MLP, taking 39 s for one training epoch, compared to 12 s for MLP. Conventional ML models are not compared on efficiency as they are running on CPU.

\vspace{-2em}
\begin{table}[!htbp]
\centering
\caption{Accuracy (\%) comparison between models w/o or w/ label hierarchy}
\label{table:accuracy_comparison}
\scalebox{0.9}{
\begin{tabular}{cccccccccccc} 
\toprule
                          &              & \multicolumn{2}{c}{AdaBoost} & \multicolumn{2}{c}{kNN} & \multicolumn{2}{c}{SVM} & \multicolumn{2}{c}{MLP} & \multicolumn{2}{c}{H-HAR}  \\
Dataset                   & Classifier   & w/o H. & w/ H.               & w/o H. & w/ H.          & w/o H. & w/ H.          & w/o H. & w/ H.          & w/o H. & w H.              \\ 
\midrule
\multirow{2}{*}{DaLiAc}   & single-label & 80.0   & 86.64               & 68.71  & 85.48          & 54.13  & 87.12          & 88.92  & 94.62          & 91.64  & \textbf{97.43}    \\
                          & multi-label  & 76.28  & 83.34               & 64.53  & 76.32          & 52.34  & 82.34          & 88.32  & 92.43          & 90.98  & \textbf{97.23}    \\ 
\midrule
\multirow{2}{*}{UCI HAPT} & single-label & 88.96  & 92.39               & 75.62  & 88.92          & 89.26  & 94.25          & 90.54  & 96.77          & 95.45  & \textbf{97.98}    \\
                          & multi-label  & 84.23  & 89.23               & 72.43  & 84.34          & 87.23  & 92.34          & 90.23  & 95.88          & 94.32  & \textbf{97.82}    \\
\bottomrule
\end{tabular}
}
\end{table}
\vspace{-2em}

\subsection{Ablation study}
To understand why our model performs effectively, we conduct ablation studies on various parameters that might impact or enhance the model's performance. Specifically, as detailed in Table~\ref{table:ablation_study}, we examine several H-HAR variants:
\begin{itemize}
    \item Label Hierarchy
    \begin{itemize}
        \item None: replace the graph modeling layer in Equation~\ref{eq:predefined_learnable_graph} with a linear layer;
        \item $\hat{A}$: only use the predefined label hierarchy for label modeling;
        \item $\hat{A}_{adp}$: only employ a learnable graph-based label modeling.
    \end{itemize}
    \item Feature Propagation (None): replace Feature Propagation by a linear layer
    \item Objective Function
    \begin{itemize}
        \item $L_{align}$+$L_{ce}$: label-data alignment loss with cross-entropy loss;
        \item $L_{con}$+$L_{ce}$: contrastive loss with cross-entropy loss;
        \item $L_{ce}$: only cross-entropy loss.
    \end{itemize}
\end{itemize}

\begin{table}
\centering
\vspace{-2em}
\caption{Ablation study: model accuracy (\%) w.r.t. various parameters}
\label{table:ablation_study}
\scalebox{0.85}{
\begin{tabular}{ccp{1.1cm}p{1cm}p{1cm}ccccc} 
\toprule
                          &              & \multicolumn{3}{c}{Label Hierarchy} & Feat. Propag.& \multicolumn{3}{c}{Objective Function}               & \multirow{2}{*}{H-HAR}  \\
Dataset                   & Classifier   & None  & $\hat{A}$ & $\hat{A}_{adp}$ & None                & $L_{align}$+$L_{ce}$ & $L_{con}$+$L_{ce}$ & $L_{ce}$ &                         \\ 
\midrule
\multirow{2}{*}{DaLiAc}   & single-label & 91.64 & 94.23     & \textbf{97.69}           & 96.23               & 94.32                & 97.33              & 94.42    & 97.43          \\
                          & multi-label  & 90.98 & 94.12     & 97.21           & 95.67               & 93.23                & 96.59              & 92.38    & \textbf{97.23}          \\ 
\midrule
\multirow{2}{*}{UCI HAPT} & single-label & 95.45 & 97.32     & \textbf{98.20}           & 97.45               & 97.28                & 97.89              & 96.73    & 97.98          \\
                          & multi-label  & 94.32 & 97.24     & 97.12           & 97.12               & 96.52                & 97.65              & 95.72    & \textbf{97.82}          \\
\bottomrule
\end{tabular}
}
\end{table}

From the results, we observe that i) The model performs better in single-label classification with just the learnable graph than when combined with a predefined label hierarchy. This suggests that learning relationships directly from data can be more effective than using pre-set connections (\textbf{RQ 2}); ii) Adding feature propagation improves the model's performance. This likely happens because it helps align data better with the graph's structure; iii) The biggest boost in performance comes from supervised contrastive learning, which helps build class-separable embeddings. Joint optimization of these techniques also helps enhance the model's overall effectiveness (\textbf{RQ 3}).
\section{Discussions and Conclusion}
\label{sec:conclusion}
Modeling and integrating label relationships into HAR models allows regularizing the representation space, thus building better feature embeddings. The proposed H-HAR brings multiple research opportunities, which are not fully addressed in the paper: i) the hierarchy-aware label modeling allows us to handle data with heterogeneous-granular labels, leading to less effort and better flexibility in practice for data annotations; ii) the contrastive learning can be further explored in the context of label relationship modeling. For instance, a hierarchy-aware margin parameter can be investigated~\cite{chen2021hierarchy}; etc.

\noindent \textbf{Conclusion} In this work, we propose H-HAR and rethink Human Activity Recognition (HAR) tasks from a perspective of graph-based label modeling. The proposed hierarchy-ware label encoding can be seamlessly integrated into other HAR models to improve further models' performance. For future work, one can be exploring more complex data with a deeper hierarchy and intricate label relationships. Human activities with multi-modality will also be one of the research directions in the future.
%



\bibliographystyle{splncs04}
\bibliography{reference.bib} 
\end{document}